\documentclass{osa-article}

\journal{boe}
\usepackage{placeins}
\usepackage{hyperref}
\articletype{Research Article}
\begin{document}

\title{Scattering oblique plane microscopy for \textit{in-vivo} blood cell imaging}

\author{Gregory N. McKay,\authormark{1} Ryan C. Niemeier,\authormark{1} Carlos Castro-Gonz\'alez,  \authormark{2} and Nicholas J. Durr\authormark{1,*}}

\address{\authormark{1}Department of Biomedical Engineering, Department of Electrical and Computer Engineering, Johns Hopkins University\\
\authormark{2}Leuko Labs Inc., Boston, MA, USA\\}

% \authormark{2}Publications Department, The Optical Society (OSA), 2010 Massachusetts Avenue NW, Washington, DC 20036, USA\\
% \authormark{3}Currently with the Department of Electronic Journals, The Optical Society (OSA), 2010 Massachusetts Avenue NW, Washington, DC 20036, USA}

\email{\authormark{*}ndurr@jhu.edu} %% email address is required

\homepage{http://durr.jhu.edu/} %% author's URL, if desired


\begin{abstract} 
Oblique plane microscopy (OPM) enables high speed, volumetric fluorescence imaging through a single-objective geometry. While these advantages have positioned OPM as a valuable tool to probe biological questions in animal models, its potential for \textit{in vivo} human imaging is largely unexplored due to its typical use with exogenous fluorescent dyes. Here we introduce a scattering-contrast oblique plane microscope (sOPM) and demonstrate label-free imaging of blood cells flowing through human capillaries \textit{in vivo}. The sOPM illuminates a capillary bed in the ventral tongue with an oblique light sheet, and images side- and back- scattered signal from blood cells. By synchronizing sOPM with a conventional capillaroscope, we acquire paired widefield and axial images of blood cells flowing through a capillary loop. The widefield capillaroscope image provides absorption contrast and confirms the presence of red blood cells (RBCs), while the sOPM image may aid in determining whether optical absorption gaps (OAGs) between RBCs have cellular or acellular composition. Further, we demonstrate consequential differences between fluorescence and scattering versions of OPM by imaging the same polystyrene beads sequentially with each technique. Lastly, we substantiate \textit{in vivo} observations by imaging isolated red blood cells, white blood cells, and platelets \textit{in vitro} using 3D agar phantoms. These results demonstrate a promising new avenue towards \textit{in vivo} blood analysis.
\end{abstract}

%%%%%%%%%%%%%%%%%%%%%%%%%%  body  %%%%%%%%%%%%%%%%%%%%%%%%%%
\section{Introduction}
The recent application of modern microscopy techniques to human capillary imaging promises non-invasive alternatives to conventional, phlebotomy-based clinical blood tests. For example, neutropenia screening has recently been demonstrated non-invasively by imaging nailfold capillaries, relying solely on intrinsic absorption-based contrast produced by hemoglobin within red blood cells \cite{Bourquard2018,Bathini2019}. Capillaroscopy holds the potential to enable safer, at-home screening for immunocompromised patients. This includes cancer patients who are at increased risk of infection from the side effects of chemotherapy, and who have infrequent access to phlebotomy-based blood cell counts in hospital settings. Low-cost, mobile phone-based versions of this technology have been developed, indicating this technique could have further impact as a point-of-care tool in low- and middle- income countries that lack access to laboratory equipment \cite{Castro-Gonzalez2018,McKay2020b}. Neutropenia screening with nailfold capillaroscopy relies on detecting and quantifying optical absorption gaps (OAGs) that are most likely to contain a neutrophil, based on features described in literature \cite{Schmid-Schonbein1980,Sinclair2018,Roggan1999,Fuchsjager-Mayrl2002,Uji2012,Golan2012a}. OAGs appear as gaps of negative absorption contrast between red blood cells within a capillary loop, and there is strong evidence they can be caused by circulating leukocytes or blood plasma \cite{Schmid-Schonbein1980,Sinclair2018,Roggan1999,Uji2012}. Currently, there is no method of determining the composition of an OAG. To produce quantitative white blood cell counts non-invasively, there is a need for tools that elucidate the composition of OAGs flowing through human capillaries.

Techniques such as optical coherence tomography \cite{Ginner2018, Chen2017, Leitgeb2019} and reflectance confocal imaging \cite{Saknite2020} could be applied to \textit{in vivo} cytometry, however compared to OPM, they are typically lower resolution and lower speed, respectively. Spectrally encoded confocal microscopy \cite{Tearney1998,Golan2010,Golan2012a} can increase confocal imaging speed but requires additional system complexity. A conventional scattering light sheet microscope with two orthogonal objectives can also probe scattering signal in bulk tissue \cite{Nguyen2020}, however is not compatible simultaneously with widefield capillaroscopy as presented here. Oblique back-illumination capillaroscopy \cite{McKay2020a}, a relatively simple, widefield imaging technique, could prove useful in resolving OAGs in superficial capillaries, though only images a single depth.

For the task of detecting blood cells flowing in capillaries \textit{in vivo}, and in particular detecting white blood cells, scattering-contrast oblique plane microscopy (sOPM) provides several advantageous features. From a microscopy perspective, human capillaries are superficially embedded within a turbid, semi-infinite medium, and thus, a single-objective, epi-illumination geometry is ideal \cite{Dunsby2009, Booth2008}. Further, an oblique light sheet probes a range of tissue depths simultaneously with an en face view, which is useful for finding and monitoring capillaries embedded within tissue at different depths. sOPM can utilize low-cost high-speed cameras to meet the frame-rate demands of capturing blood cells that travel up to 1mm/s at the capillary level  \cite{Bollinger1974,Mugii2009}. Though it broadens with depth in scattering media such as skin, the oblique light sheet provides optical sectioning, illuminating primarily a thin plane of tissue conjugate with the sensor, helping reduce out-of-focus signal. A well-formed light sheet can illuminate blood cells passing single-file through a capillary, analogous to a conventional scattering-based flow cytometer \cite{Henel2007}. In fact, Dunsby suggested the use of OPM for flow cytometry in his seminal work \cite{Dunsby2009}. In conventional flow cytometry, the intrinsic side- and forward-scattering signal from laser-illuminated blood cells provides identifying information on cellular granularity and size \cite{Cho2010,Ost1998}. In contrast, sOPM detects side- and back-scattered light. 

In this paper, we build an OPM and compare fluorescence and scattering contrast by imaging the same polystyrene bead sample with each technique. We also use sOPM to image the intrinsic scattering signal from blood cells flowing in a human capillary \textit{in vivo}. By simultaneously acquiring data of the same capillary with sOPM and a conventional widefield capillaroscope, we demonstrate that OAGs that appear to be comprised of a white blood cell can be distinguished with sOPM from those that appear to be plasma gaps. Finally, we confirm a similar scattering measurement \textit{in vitro} from isolated blood cells embedded in a 3D agar phantom. 

\section{Methods}

\subsection{Optical system}
\begin{figure}[h!] 
\centering\includegraphics[width=14cm]{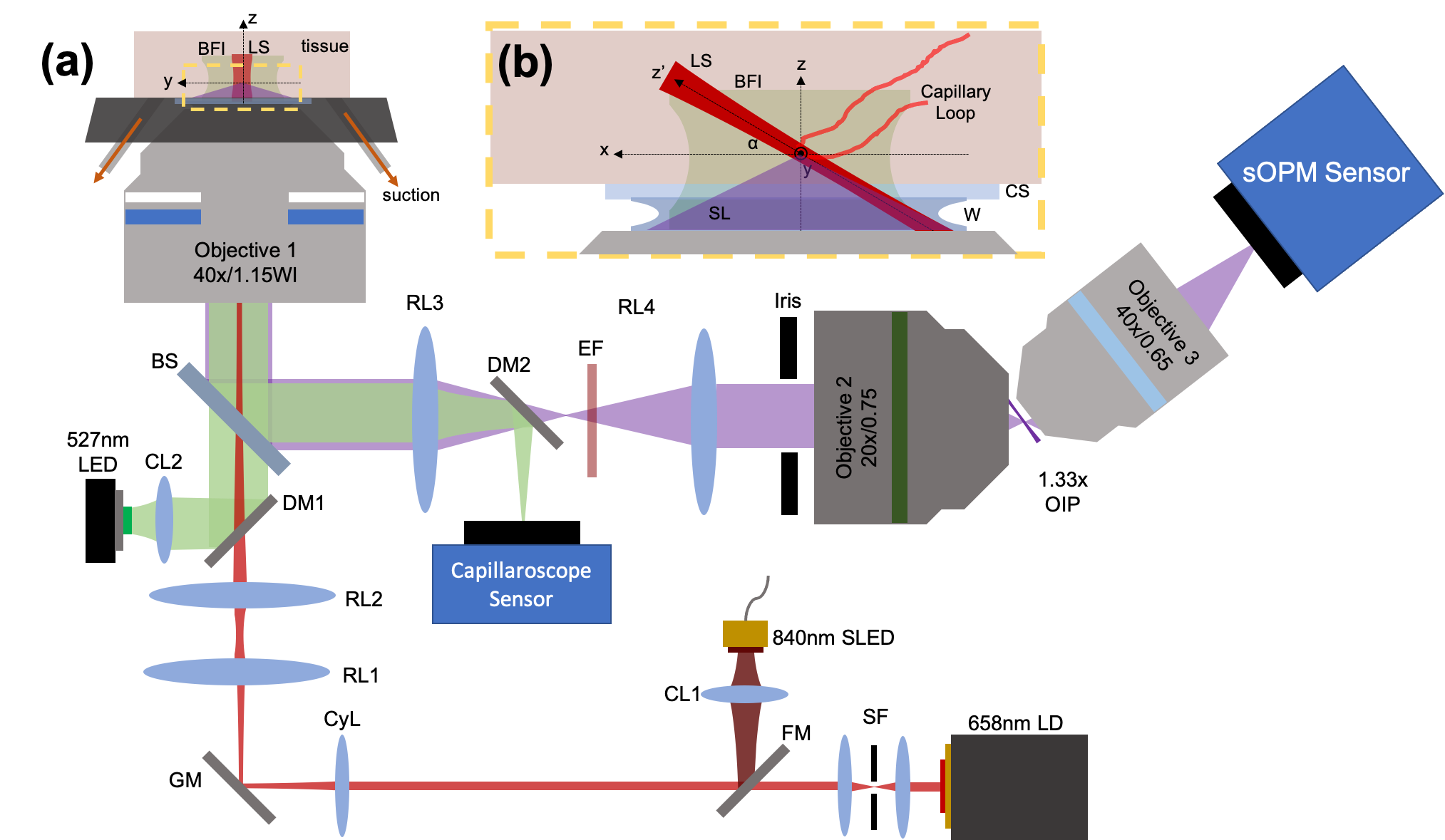}
\caption{(a) Layout of the sOPM (red and purple beamlines) and the widefield capillaroscope (green beamlines). A flip mirror (FM) is used to alternate between 658nm laser diode (LD) or 840nm super luminescent diode (SLD) light sheet (LS) illumination. The LD is shaped by a spatial filter (SF), and the 840nm diode is collimated with a collimating lens (CL1). Both sources pass through a cylindrical lens (CyL), which forms a light sheet onto galvanometric mirrors (GM), which is then imaged to the back aperture of objective 1 by relay lenses (RL) 1 $\&$ 2.  Scattered light (SL) is collected by the same objective, reflected off a beamsplitter (BS), and relayed by another 4F system (RL3 and RL4) to the back aperture of objective 2, forming a 1.33x magnification oblique image plane (OIP). Finally, a finite conjugate objective 3 images the oblique image plane onto the sOPM sensor. For fluorescence-mode imaging, an emission filter (EF) is placed prior to objective 2. For aberration control, a circular iris is placed at the back aperture of objective 2. Widefield capillaroscopy with brightfield illumination (BFI) is performed with a collimated 527nm LED introduced through a dichroic mirror (DM1). Light is collected by the same objective and directed to the capillaroscope sensor by a second dichroic mirror (DM2). (b) An enlarged region of interest of the sample space shows a capillary loop within the light sheet and the imaging coordinate system. Note that z' refers to the direction of the light sheet. Scattered light is detected through a coverslip (CS) and water immersion (W).}
\label{fig:Figure1}
\end{figure}

A schematic of the optical system used for the experiments in this paper is shown in Figure \ref{fig:Figure1}. The system is comprised of two main components: (1) the sOPM, and (2) a widefield capillaroscope. 

The sOPM system is illuminated by either a super luminescent diode (SLD, 840nm Exalos EBD270001-03) collimated with a 15 mm biconvex lens, or a laser diode (658nm, Thorlabs L658P040) collimated and spatially filtered (SF) with a 50 $\mu$m pinhole placed at the Fourier plane of a 2x beam expander (25 mm and 50 mm biconvex lenses). Sources are switchable using a flip mirror (FM) in the common illumination path, and are focused to a line on one mirror of a galvanometer mirror pair (GM, Thorlabs GVS012) by a cylindrical lens (f = 50 mm). The line on the mirror is made conjugate with the back focal plane of objective 1 (Nikon 40x 1.15NA APO LWD WI $\lambda$S WD = 0.59-0.61) by a 4F system (RL1, f = 75mm and RL2, f = 150mm). The image of the light sheet at the back focal plane is offset from the objective's optic axis, creating an oblique light sheet (LS) in sample space. The GMs were rotated during alignment to tune the light sheet angle to $\alpha$ = 36$^{\circ}$. The system is aligned such that the working distance of objective 1 is located at the sample surface. Scattered light (SL, purple) is collected by objective 1, redirected off a 50:50 beamsplitter (BS, Thorlabs BSW29), and relayed to objective 2 (Nikon 20x 0.75NA Plan Apo DIC N2 WD = 1mm) by a second relay system (RL3, f = 100mm, RL4, f = 150mm). A circular iris (Thorlabs, SM1D12C) is placed at the back aperture of objective 2 to reduce its effective numerical aperture and control aberrations. The collected light is imaged to an oblique image plane (OIP) with net magnification from the sample space of 1.33x, to match the index of refraction ratio between objective 1 and objective 2 sample space. Finally, objective 3 (Nikon finite-conjugate 40x 0.65NA 210 ELWD) re-images the tilted light sheet on the sOPM sensor (PCO.edge 5.5 sCMOS). For fluorescent bead imaging, a 680nm LP filter (EF, Omega RPE680LP) is added in the second 4F system. The magnification of the sOPM system is 53.2x, considering the index of refraction 1.33 and the 40x magnification of objective 3. 

A widefield capillaroscope system is included to enable simultaneous acquisition of the two different signals. It consists of a green LED (Superbright 1W XLamp LED, 88lm, 527nm) condensed by an aspheric lens (f = 20mm, Thorlabs ACL2520U-A), and reflected off a 605nm long-pass dichroic mirror (DM1, Thorlabs DMLP605R). The LED produces brightfield illumination (BFI) on the sample, that is strongly absorbed by hemoglobin within red blood cells. Remitted light is collected by objective 1, redirected off the beamsplitter, and focused onto the capillaroscope sensor (Imaging Source DMK 33UX252) by RL3 and a second dichroic mirror (DM2, Thorlabs DMLP605R). The magnification of the capillaroscope is approximately 20x, due to RL3, acting as tube lens with half the standard focal length specified for objective 1.

\FloatBarrier

\subsection{Polystyrene bead phantoms}
Fluorescent polystyrene beads (4$\mu$m diameter, 1$\%$ solids, Bangslabs FSFR006) were embedded in 1$\%$ w/v aqueous agar solution with a 30x dilution. Bead solution was pipetted into 800$\mu$m deep silicone isolators (Invitrogen secure-seal hybridization chamber gasket), which was mounted on top of a coverslip ($\#$1.5) for imaging. The 658nm laser diode was used for illumination, and an exposure time of 50ms was used for scattering imaging. Since the scattering signal was significantly more intense than fluorescence, the image was dominated by scattered light in this case. For fluorescent imaging, an emission filter (Omega RPE680LP) was inserted into the detection path, blocking the scattering signal, and an exposure time of 500ms was used.

\subsection{\textit{In vivo} capillary imaging}
\textit{In vivo} human imaging was conducted in a protocol approved by the Johns Hopkins University Institutional Review Board (IRB00204985). A single healthy, human volunteer was enrolled in the study. The participant's ventral tongue was imaged at 100Hz with the 658nm laser diode, using vacuum stabilization to keep the tissue stationary (see \cite{McKay2020a} for details). A digital hardware trigger was used (Arduino Uno) to synchronize image acquisition between the sOPM and capillaroscope sensors.

\subsection{Blood cell phantoms}
\begin{figure}[h!] 
\centering\includegraphics[width=12cm]{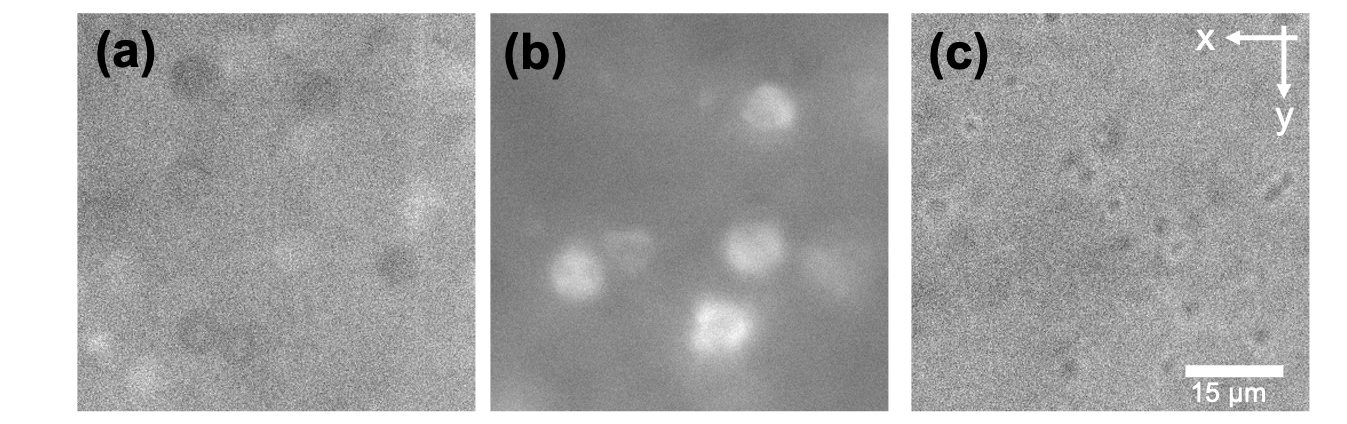}
\caption{Purified blood cell products were obtained and confirmed with the capillaroscope - (a) leukocyte-depleted red blood cells, (b) purified white blood cells from buffy coat, and (c) concentrated platelets.}
\label{fig:Figure2}
\end{figure}

To support the \textit{in vivo} blood cell scattering results, we created \textit{in vitro} blood cell phantoms using 1.5 $\%$ w/v low melting point (LMP) agar solution in a 0.9 mM NaCl buffer solution. Blood products were purchased from Zen-Bio, including leukocyte-depleted erythrocytes (SER-LDRBC-SDS), buffy coat (SER-BC-SDS), and platelet-rich concentrate (SER-PC-SDS). White blood cell isolation was conducted on the buffy coat sample, using centrifugation (300g, 5 minutes) and AKC red blood cell lysis buffer (ThermoFisher A1049201). The platelet sample was further concentrated 8x through centrifugation (1500g, 15 minutes), pelleting, decanting, and resuspension. Purified blood products were diluted 20x with LMP agar solution, and pipetted into 800 $\mu$m deep silicone isolators. Images of the purified products were taken with the widefield capillaroscope to verify successful cell purification (Figure \ref{fig:Figure2}).

\FloatBarrier

\section{Results}

\subsection{Polystyrene bead phantoms}
\begin{figure}[h!] 
\centering\includegraphics[width=12cm]{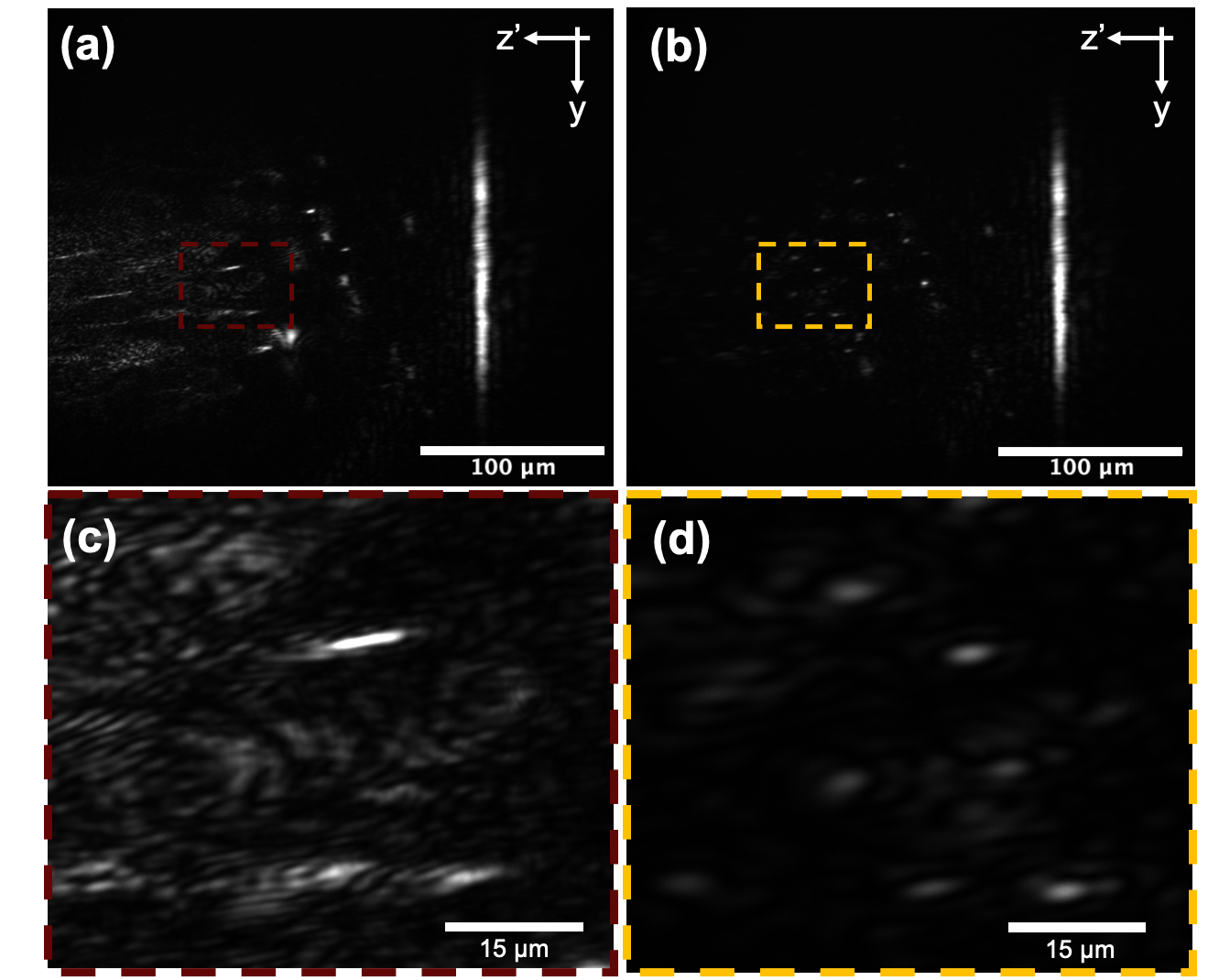}
\caption{(a) Aberration and depth of field control demonstrated with 4 $\mu$m polystyrene beads and a variable iris placed at the back focal plane of objective 2. Full frame (a) and region of interest (c) taken with objective 2 NA $= 0.75$. Full frame (b) and region of interest (d) taken with objective 2 NA$_{eff} = 0.33$. An increase in particle number, likely due to depth of field increase, and improved particle visibility, from aberration control, are both observed by decreasing the effective numerical aperture of collection. The vertical line of intensity superficial to the sample is formed by specular reflection from the coverslip.}
\label{fig:Figure3}
\end{figure}

Polystyrene bead phantoms provided insight into a few key phenomena. First, we observed that their image appeared to become severely aberrated at sample depths of z > 95 $\mu$m (z' > 160 $\mu$m) (Figure \ref{fig:Figure3}(a) $\&$ (c)). This is consistent with the axial range of aberration-free remote focusing predicted by Botcherby et al.'s analysis adapted to our system parameters \cite{Booth2008}. We observed that the aberrations at these imaging depths could be reduced with an iris at the back aperture of objective 2 (Figure \ref{fig:Figure3}(b) $\&$ (d)), which reduces the effective numerical aperture of the objective. A 4 mm diameter iris, which creates an effective numerical aperture for objective 2 of NA$_{eff} = 0.33$, provided a reasonable balance of imaging depth and bead resolution. Additionally, we observed that more polystyrene beads came into focus when we stopped down the iris (Figure \ref{fig:Figure3}(c) vs. (d)). We believe this is due to an increase in depth of field afforded by the lower numerical aperture, now focusing light from particles located in planes slightly above and below the center axis of the light sheet. While this technique does improve particle resolution by controlling aberrations, the collected scattering signal was significantly reduced (Figure \ref{fig:Figure3}(c) vs. (d)).

\FloatBarrier

\begin{figure}[h!] 
\centering\includegraphics[width=12cm]{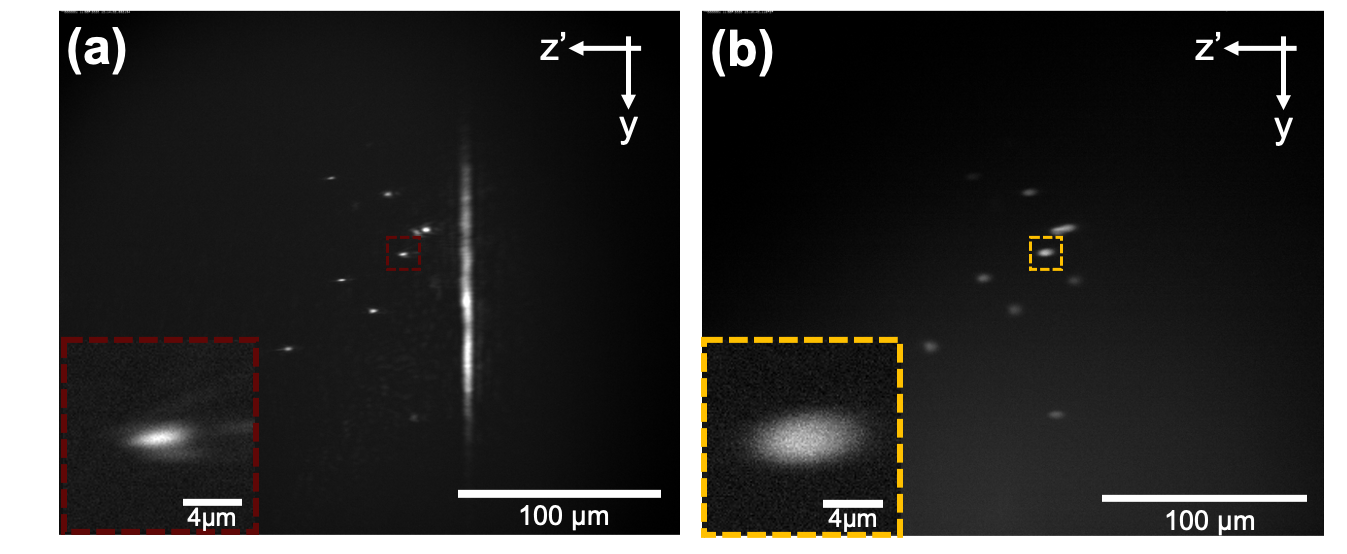}
\caption{(a) Scattering- and (b) fluorescence-OPM images of the same 4 $\mu$m polystyrene beads in agar. The same beads imaged with scattering contrast appear nearly half the bead size, likely a virtual image from reflected light off the spherical bead.}
\label{fig:Figure4}
\end{figure}

We also observed that polystyrene beads imaged by the sOPM system appeared consistently smaller than what was predicted by the magnification and the 4 $\mu$m specified bead diameter (Figure \ref{fig:Figure4}(a)). In scattering detection, a FWHM of 1.1 $\mu$m was observed along the y-axis, and 3 $\mu$m along the z' axis. However, when the same beads were imaged by only their fluorescent signal by placing a 680nm long-pass emission filter in the detection path and increasing the exposure time, the beads appeared much closer to their expected size (Figure \ref{fig:Figure4}(b)). With fluorescent detection, a FWHM of 3.5 $\mu$m was observed along the y-axis, and 5.9 $\mu$m along the z' axis. We believe the discrepancy between the two signals is due to higher order scattering artifact (lensing and mirroring) arising from the large index of refraction mismatch between the polystyrene beads and the agar phantom \cite{Ledwig2021}. The small, spherically shaped beads may act as miniature convex mirrors, creating a demagnified virtual image within the bead itself. In the case of fluorescent imaging, specular reflection is blocked by the emission filter, and the signal of the entire fluorescent bead is observed.

\FloatBarrier
\subsection{\textit{In vivo} capillary imaging}

\begin{figure}[h!] 
\centering\includegraphics[width=12cm]{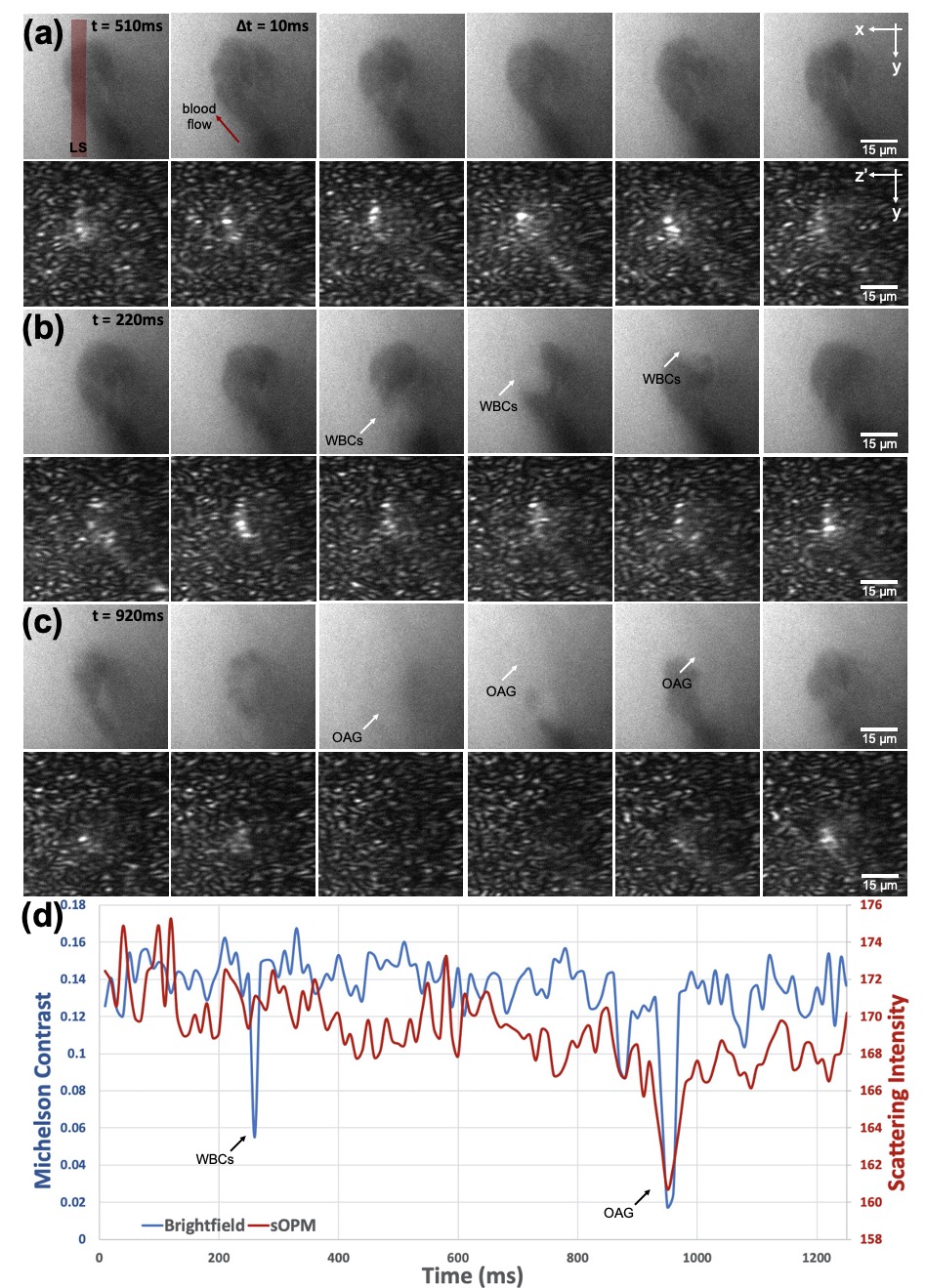}
\caption{Paired widefield capillaroscopy and sOPM images of an \textit{in vivo} human ventral tongue capillary. Each column represents 10 ms time difference, light sheet (LS) location denoted by red overlay. (a) Red blood cells pass through the capillary loop as dark shadows in widefield capillaroscopy imaging (top), and at the same time generate strong scattering signal in sOPM image (bottom). (b) A tightly packed pair of white blood cells (WBCs) passes around the capillary loop and generates a similarly strong scattering signal to adjacent red blood cells in sOPM imaging. (c) A large OAG passes through the capillary loop, producing a marked decrease in scattering intensity observed by the sOPM sensor, suggesting acellular origin. (d) The Michelson contrast of the capillary in the capillaroscope image is compared to the mean scattering intensity from sOPM.}
\label{fig:Figure5}
\end{figure}

We next imaged capillaries on the ventral tongue of a healthy human participant. We used vacuum stabilization to keep the tongue stationary \cite{McKay2020a} and imaged with both scattering-contrast OPM and absorption contrast widefield capillaroscopy. A time-course measurement from this experiment is displayed in Figure \ref{fig:Figure5}. Each panel (a)-(c) highlights a distinct event that occurred while imaging the same capillary, with widefield capillaroscopy images placed directly above the temporally corresponding sOPM images, and temporally adjacent frames in each column ($\Delta$t = 10 ms). In Figure \ref{fig:Figure5}(d), a quantitative comparison between line profiles of the image modalities is presented. The Michelson contrast of the capillary in the capillaroscope is compared to the mean scattering intensity of the sOPM signal. When red blood cells pass through the capillary (Figure \ref{fig:Figure5}(a)), we observe a relatively weak and consistent signal in the widefield capillaroscope (due to hemoglobin absorption) and a relatively strong and consistent signal in sOPM (due to intracellular scattering). When a pair of white blood cells pass through the capillary (Figure \ref{fig:Figure5}(b)), a distinct drop in capillary contrast is observed with the widefield system from the absence of hemoglobin. For this event, the sOPM system registers a scattering signal from the pair of white blood cells that is similar in intensity to that from the red blood cells. Finally, in Figure \ref{fig:Figure5}(c), a larger OAG is observed passing through the capillary, where the presence or absence of white blood cell(s) is more ambiguous. In this case, a wider but similar drop in contrast is observed in the widefield capillaroscope. For this event however, the sOPM system also demonstrates a marked decrease in scattering intensity, suggesting the OAG is acellular in nature and could be caused simply by blood plasma.

\FloatBarrier

\subsection{Blood cell phantoms}

\begin{figure}[h!] 
\centering\includegraphics[width=14cm]{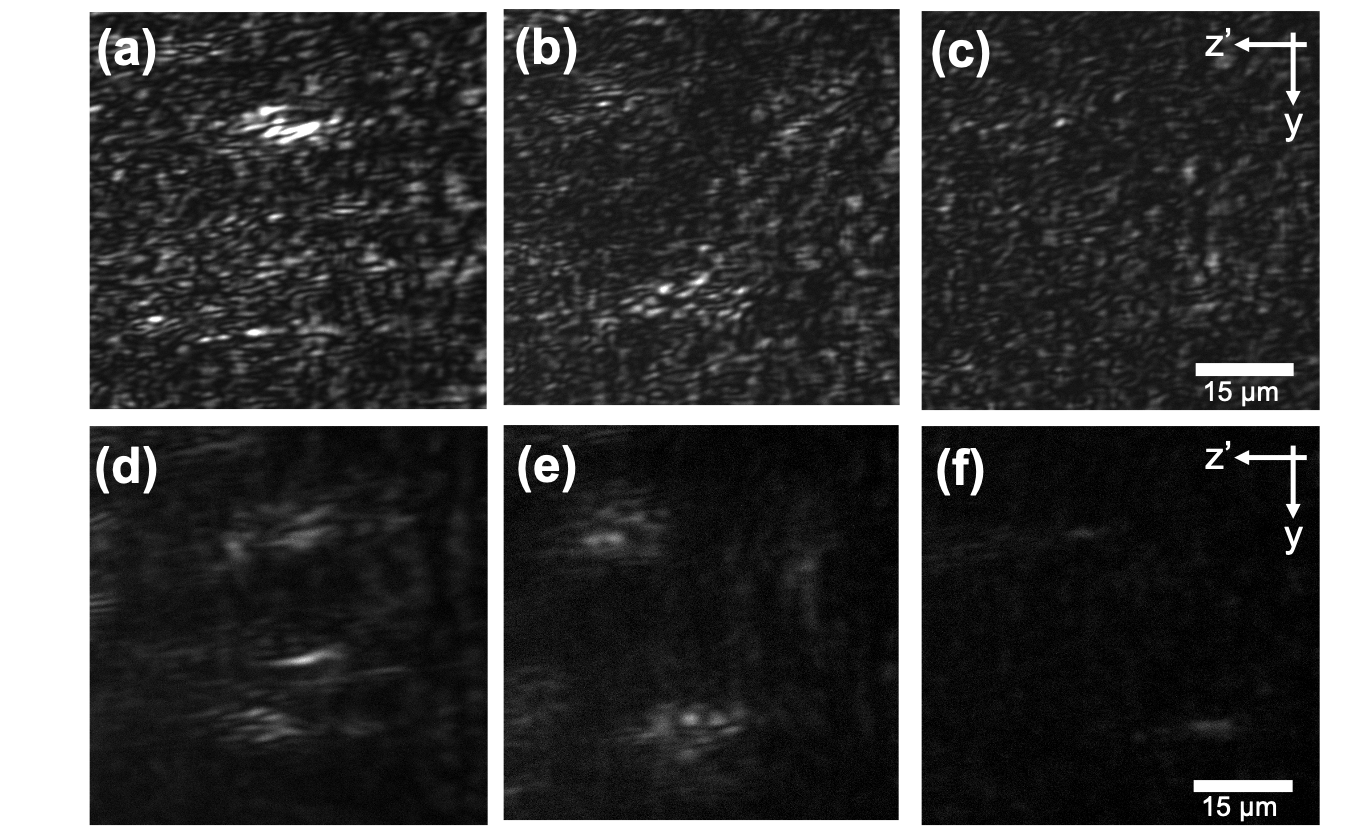}
\caption{Agar blood cell phantom imaging. Panels (a) $\&$ (d) show isolated red blood cells, (b) $\&$ (e) show white blood cells, and (c) $\&$ (f) show platelets. Each sample was first imaged with the 658nm laser diode (top), and second with an 840nm SLD (bottom).}
\label{fig:Figure6}
\end{figure}

To validate the \textit{in vivo} blood cell scattering results, we imaged isolated blood components in the sOPM system (Figure \ref{fig:Figure6}). The left column ((a) $\&$ (d)) shows RBCs, the middle column ((b) $\&$ (e)) WBCs, and the right column ((c) $\&$ (f)) show platelets. The top row ((a)-(c)) demonstrate RBC, WBC, and platelet samples, respectively, with 658nm LD illumination, while the bottom row ((d)-(f)) show the same particles illuminated with the 840nm SLD. With these experiments, we observe that red blood cells and white blood cells produce similar scattering contrast \textit{in vitro} and \textit{in vivo}. Further, we observe that the use of a SLD reduces speckle noise and makes blood cells easier to resolve. This is especially apparent with smaller particles, such as platelets, whose sizes are similar to the speckle grains of the 658nm LD. While the 840nm SLD in our experiment was not powerful enough to be used in our current system for \textit{in vivo} imaging, it is clear the use of a broadband source or other methods of speckle reduction can significantly improve the signal to noise ratio of sOPM.

\FloatBarrier

\section{Discussion}

With its epi-illumination geometry, reliance on endogenous signal, and capability for high-speed, optically-sectioned imaging, scattering oblique plane microscopy could serve as a useful tool for \textit{in vivo} blood analysis. Our results show that the endogenous scattering signal of blood cells within human capillaries can be imaged non-invasively at 100 Hz. We observe that white blood cells and red blood cells produce a similar strength of scattered signal, while larger, acellular OAGs demonstrate a marked reduction in scattered signal. While the sOPM presented here could not distinguish white blood cells and red blood cells, it can distinguish between acellular plasma gaps and blood cells. In combination with widefield absorption-based capillaroscopy, which can distinguish between red blood cells and white blood cells, the three of these major constituents of capillary blood (red blood cell, white blood cell, and plasma gaps) can be classified. Recently, the frequency of OAGs in nailfold capillaries has been inversely correlated to severe neutropenia \cite{Bourquard2018,Bathini2019}. Although the OAGs identified in this approach fulfill certain features that maximize their likelihood of representing neutrophils, they still might be either plasma or white blood cell in origin, which may limit the expansion of this technique to provide quantitative white blood cell counts. sOPM has the potential to reduce this ambiguity. However, in order to image capillaries in the nailfold, which are typically 150-400 $\mu$m beneath the tissue surface \cite{Baran2015}, improvements are needed to enable sOPM at larger depths. This may be possible with lower magnification objective lenses, longer wavelengths, and potentially incorporating confocal line-scanning to improve optical sectioning \cite{Rajadhyaksha1999a,Larson2011,Glaser2016b}. It may also be useful to test sOPM with an anamorphic lens design \cite{Shao2020} or utilizing a diffractive approach to help increase the field of view \cite{Udkewitz2019}.

Additional future work should focus on developing scattering contrast techniques that distinguish between red blood cells and white blood cells. We believe it is worth investigating scattering spectral dependence, polarization-sensitivity, and scattering angle-resolved detection for this purpose. Additionally, it is clear from our \textit{in vitro} blood cell phantom results that speckle noise is degrading the signal. Thus, the use of a higher power broadband source or other means of speckle reduction such as a commercial laser speckle reducer, rotating diffuser, or despeckling algorithms \cite{Bobrow2018} are worth investigating. Finally, volumetric tissue imaging can be achieved by incorporating scanning and descanning galvanometric mirrors into the sOPM design, similar to how SCAPE is used for fluorescence imaging \cite{Bouchard2015a}. With improved designs, scattering oblique plane microscopy may find use in dermatologic, ophthalmic, and endoscopic settings, similar to optical coherence tomography, for evaluating bulk tissue non-invasively with B-mode imaging. The ability for sOPM to detect side- and back-scattered light could enable further cell and tissue discrimination, analogous to how conventional flow cytometry extracts information through measuring side- and forward-scattered light.

\section{Conclusion}

Human capillary imaging with new microscopy techniques has the potential to reveal important clinical biomarkers that may enable non-invasive blood analysis. For example, non-invasive severe neutropenia screening has been shown possible by quantifying certain OAGs in conventional capillaroscopy videos \cite{Bourquard2018,Bathini2019}. However, despite its success to screen for neutropenia, the exact composition of OAGs is ambiguous with widefield capillaroscopy, which may limit its application to provide quantitative white blood cell count. Scattering oblique plane microscopy can provide additional and complimentary signal to discriminate cellular versus acellular OAGs in human capillaries \textit{in vivo}, helping to classify blood columns into the three major components of red blood cells, white blood cells, and plasma.

\section*{Funding}
This research was supported in part with a gift from Fifth Generation, Inc., the Johns Hopkins Medical Science Training Program Fellowship, and the National Institutes of Health (R44CA228920). 

\section*{Acknowledgments}
The authors would like to thank Milind Rajadhyaksha and Charles A. DiMarzio for helping interpret scattering vs. fluorescent signals. The authors would also like to thank Ian Butterworth, Aur\'elien Bourquard, \'Alvaro S\'anchez-Ferro and the team at Leuko Labs for their ongoing insight, support, and feedback about the project.

\section*{Disclosures}
The authors are co-inventors on a provisional patent application assigned to Johns Hopkins University. They may be entitled to future royalties from intellectual property related to the technologies described in this article.

\bibliography{sample}

\end{document}